# Using Social Network Service to determine the Initial User Requirements for Small Software Businesses

Nazakat Ali[1], Jang-Eui Hong[1]

[1] *Department of Computer Science, Chungbuk National University*
*Chungdarro 1, Seonwongu, Cheongju, 28644, Rep. of Korea*
nazakatali@selab.cbnu.ac.kr[1]

## Abstract

**Background/Objectives**: Software engineering community has been studied extensively on large-sized software organizations and has provided suitable and interesting solutions. However, small software companies that make a large part of the software industry have been overlooked.

**Methods/Statistical analysis**: The current requirement engineering practices are not suitable for small software companies. We propose a social network-based requirement engineering approach that will complement the traditional requirement engineering approaches and will make it suitable for small software companies.

**Findings**: We have applied our SNS-based requirements determination approach to knowing about its validity. As a result, we concluded that 33.06 % of invited end-users participated in our approach and figured out 156 distinct user requirements. It has been seen that it was not necessary for users to have requirements engineering knowledge to participate in our proposed SNS-based approach that made maximum users to be involved during requirements elicitation process. By investigating the ideas and opinions communicated by users, we were able to figure out a high number of user requirements. It was observed that maximum user-requirements were determined within a short period of time (7days). Our experience with SNS-based approach also says that end-users hardly know about non-functional requirements and express it explicitly.

**Improvements/Applications**: we believe that researchers will consider SNS other than Facebook that would allow applying our SNS-based approach for requirements identification. We have experienced our approach with Facebook but we do not know how our approach would actually work with other SNSs.

## 1. Introduction

A significant number of the issues in software development can be associated back to software requirement engineering (RE).[1] It is observed that RE process plays a critical role in delivering quality software. The relationship between requirements engineering and software quality is obvious because software quality is defined as the capability of a software product to conform to users' requirements. Therefore, RE is the foundation for any software development. The software engineering community has paid considerable attention to the RE for large software organizations, but the RE for small software organizations are overlooked.[2] The software companies that have ten or less than ten engineers are classified into the small software companies [3], which make a lot of contribution toward the growth of software industry. In Chile, around 44% of software companies are categorized as small software companies [4], in the USA the ratio of small software companies is 78% while in Canada it is about 50% of software companies.[3] The small software companies are a big source of technological innovation, and they fill the gap that is either unexplored or profitless for big software organizations.

Despite this situation, the needs of small software companies have been overlooked by the software requirement engineering community. The small software business companies lack much research work that addresses the needs of small software businesses.[2] This may be the lack of consideration of these small businesses or false assumptions that small software companies are not different than their large-sized counterparts, that their contribution is minor in the software industry, or they do not pose any notable research challenges. Aranda J have shown several properties of small software companies that distinguish their RE from those of big organizations, and they recommended some findings and summarized the state-of-the-practice and issues of small and medium size enterprises with respect to RE.[5] According to S. Khankaew[6], the lack of customer involvement and lack of RE knowledge in the development team are among the key issues of RE in small software companies. The research[3] has investigated 24 small companies in their diagnostic study and reported that small software businesses are suffering from a number of issues like over budget, lack of customer involvement and skilled requirement engineers in software development. The authors concluded their study while giving a suggestion that a tailored RE approach is needed for small software companies because traditional RE approaches are not suitable for small software businesses. The study[7] has also discussed the challenges faced by small software companies and revealed that users are not fully involved in RE process and systems are just imposed on users. As a result, the users do not use the system properly or reject the imposed system. Due to the budget constraints, small software companies can not afford enough funds to carry out RE process[8] and start software development without considering RE which is a critical phase of software development. That is, we propose a social network-based RE approach to cope with the above problems and our proposing RE approach can fulfill the demands for short time-to-market, rapid, simple, and inexpensive approach for software development. Particularly we see its relevance within the new software paradigms like mobile computing and software ecosystems, where traditional RE approaches provide insufficient support.[9] For example, smartwatch or smartphone app developers could use our approach to elicit, negotiate, and prioritize the customer requirements.

This paper is organized as follow: Section 2 of this paper is talking about related work, and in section 3 we have elaborated our proposed approach for requirements identification from end-users using SNS. On the other hand, section 4 introduces a case study to validate our proposed approach and section 5 concludes this paper.

## 2. Related Work

Social network services (SNSs) are web-based services that permit people to (1) maintain a profile within specific system, (2) maintain a list of connected users (e.g. friends, followers), people with whom they share connections and (3) view and comment on the posts of their own or posts of connected people and vice versa.[10] SNSs are also useful because social media encourages discussions on specific issues and it is accessible to everyone. The SNSs are penetrating in proportion due to the rapid increase of a number of SNS users. It has been forecasted that the number of SNS users will reach 2.95 billion by 2020 as shown in figure 1, among all the SNSs, the Facebook remains the most popular social media around the world with 1.86 billion active users, monthly.[11]

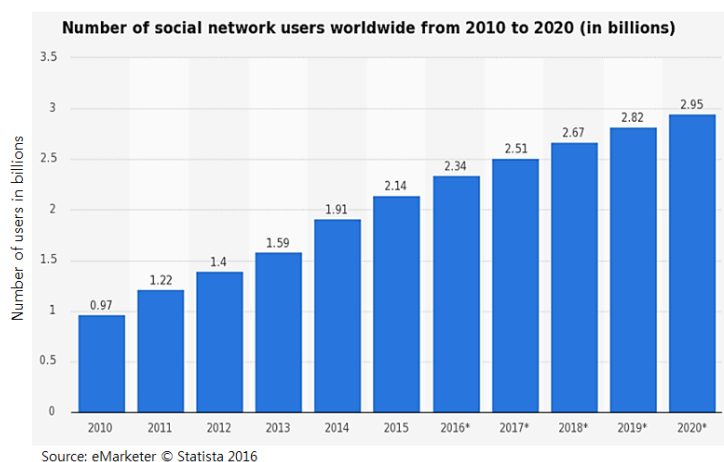

**Figure 1. SNS Users around the World**

As the number of SNS users are increasing rapidly, the SNSs have grown into an appropriate medium to collect opinions of individuals based on some data from the participants. The users of SNS usually express their experiences, and judgment without any hesitation. Therefore, so many similar opinions make a public opinion and form a useful knowledge base.[12] Researchers have proposed various requirements elicitation approaches. Some of the most famous requirements elicitation approaches are as follow.

*The interview approach*[13] is most traditional and commonly used approach for requirements elicitation, where end-users, domain experts, and stakeholders are interviewed to elicit their needs. Interviews (structured) are conducted using a predefined set of questions and it is effective to get answers that interviewer expects. The structured interviews lead to limit the opinions of customers within the framework set by the interviewers. In this approach, the target sample of customers does not represent the opinion of the majority of the customers due to temporal and spatial constraints. *Observation*[13] is also a common and widely used approach to elicit the requirements of customers. It is self-explanatory that the actual execution of the existing process is observed by users without direct involvement. Usually, this approach is used along with other approaches such as task analysis and interviews. This approach requires high skills and efforts on the side of the analyst to interpret and understand the actions being performed. It is also expensive to perform. Thus, it is not suitable for small software business because small software companies always suffering from financial constraints. *Focus Group*[14] is also a kind of requirements elicitation approach that brings groups together to discuss some topic of interest with customers. This approach is used to involve maximum potential customers during requirement elicitation. This approach is quite effective to elicit responses to products whose features and trade-offs are well known for the customers. This approach is expensive and complex for small software organizations.

In addition to above approaches, there are some other techniques to gather users' requirements such as questionnaire, and document analysis. Also, a few studies for requirements elicitation from SNS exist; N. Self[15] performed some experimental studies for requirements elicitation from SNS, but the participants of the experiments are limited to the students of the university. The university students of RE subject do not represent the common users of a system because they already have the knowledge of RE process. Y. Lee[16] proposed a technique for eliciting customer requirements based on SNS. This study is similar to our approach, but it is intended to collect a large amount of information for the large-size software business. Therefore, this technique is not appropriate for small software organization due to technical constraints, because the collected big data needs comprehensive data analysis to transform user opinions to user requirements. We propose an SNS-based RE approach that enables us to reach customers across the world to determine initial user requirements with an easy way while considering the characteristics of the small software business.

## 3. SNS-Based RE Approach

The SNS can be useful for requirements elicitation because requirements elicitation is a communications-intensive process, and can be a tool which can involve many users to improve communications, then to get fruitful user requirements. Due to some core functionalities, SNSs are valuable for software engineering community to determine initial user requirements. We can get right input from appropriate people in order to have right user requirements. Whether there is a sample of end-users or millions of end-users across the world, the SNS can allow us to hear what they have to say. With the SNS, we can not only elicit the user requirements but also closely observe all the discussions and conflicts around them that are also necessary for initial user requirements identification and prioritization.

We propose an SNS-based lightweight approach to support the RE process of small software companies. We aimed to provide an effective RE approach which is based on popular SNS (e.g. Facebook). Some popular SNSs (Facebook, Twitter) provide enough functionalities to elicit the user requirements.[17,21] Our approach supports RE activities (elicitation, negotiation, and prioritization) from an end-user point of view and needs neither a long time nor costly. Our approach is handled by a moderator, a role who has the interest to build the system, and whose goal is to gather requirements for the system. It can be used ad-hoc, anytime, and anywhere to support distributed requirements elicitation, prioritization, and negotiation in an iterative manner. The SNS-based RE approach has three phases such as preparation, requirements elicitation, and requirements determination as shown in figure 2. Our approach needs minimum training for moderators (for example, smartwatch app developer who would like to gather user requirements). A moderator is one (a person in charge of the small software company or a team member) who controls all the activities of each phase.

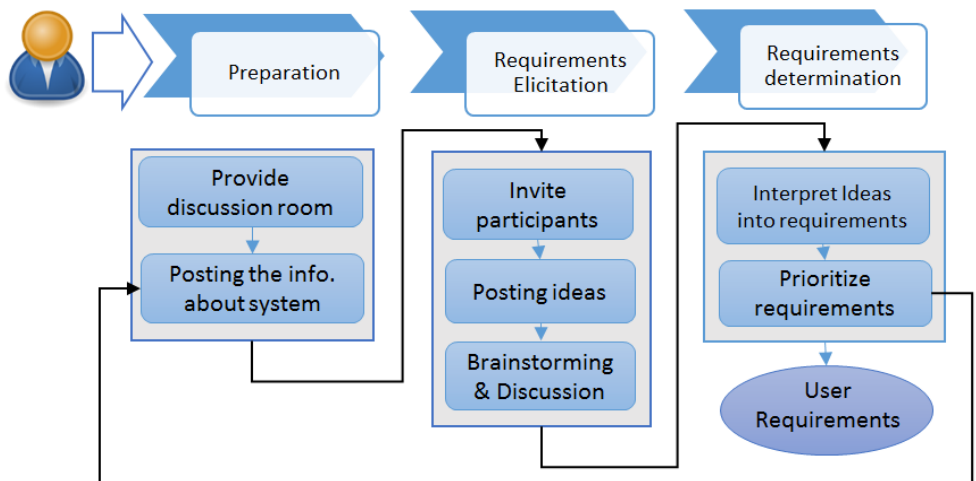

**Figure 2. SNS Based RE Approach for Small Software Companies**

### 3.1. *Preparation*

This step is mainly done by a moderator. The moderator selects an appropriate SNS (e.g. Facebook) and then creates discussion room with an appropriate title. The moderator then invites users to join the discussion room. Before starting the discussion, users might know the purpose of the discussion group and brief information about the system. The selected SNS should provide following key functionalities to initiate the RE process.

- The SNS must provide a space(room) for discussion
- The SNS must facilitate users to post their ideas
- The users should comment on the ideas
- The SNS must provide functionality to control group access

The moderator can make discussion group in private or public manner and then provides initial information (vision statement of the project) about the project. This activity should be well organized and should be viable to talk over many topics e.g. information about the project, and instruction for the participants.

### 3.2. *Requirements Elicitation*

This activity implements requirements gathering, elicitation, and negotiation. This is mainly done by participants (end users) with the guidance of a moderator.

In this activity, moderator invites users to the created group (i.e., discussion room). The invitation includes the necessary information about project and purpose of the group. Then, the moderator invites users to start posting their ideas, needs, and interests that later be converted into user requirements of the project. When the opinions are posted, users can start a discussion. Users can make queries to clear the confusion and ambiguities. The users exchange their ideas and express their needs. The moderator should start negotiation at this level by asking questions for further clarification. During the negotiations, the moderator should be engaged with participants to facilitate and lead the negotiations. In negotiations, any requirements can be criticized and can be adjusted based on the feedback of end-users. Lastly, the participants approve the important ideas by commenting on the posts. The needs of users which has high consensus are highlighted for requirements prioritization. All these three tasks of this activity can be run concurrently and in an iterative way. It is not necessary that users follow the strict sequence of these activities.

### 3.3. *Requirements Determination*

This is last activity that implements requirements determination. This phase is performed by a moderator and it deeply relies on the skills and expertise of moderator. The end users' participation may vary in quantity, quality, and content. However, the moderator is available throughout the discussion; we do not believe the high-quality description of requirements as an output, rather the identification of initial user requirements. After that, the moderator can transcribe initial user needs into user requirements. The moderator analyzes the number of approvals per idea to define the priority of requirements. The requirements prioritization will also be done based on goal, importance, and difficulty of requirements. The moderator can use some rules to prioritize user requirements.

- Rule 1: Clustering relevant requirements
- Rule 2: Prioritizing by the distance between a cluster and the requirement topic which is decided at preparation phase
- Rule 3: Adjusting priority of the requirements and cluster when the postings or opinions are duplicated
- Rule 4: Reduce certain requirements having very low priority, low feasibility or low relevance with theme of product

When determining final requirements by a moderator, we have to consider some constraints that can influence on small software company: The first constraint is not having enough budget, thus all of the collected requirements should be prioritized because some of them can be removed. The first constraint is considered already in the third activity of our proposing process. The next is the categorization of potential users due to the competitive market. Thus, the small software company has to elicit more distinguishable feature from the collected requirements even though the feature has a little bit low priority. The third one is technical constraints. This constraint may require to analyzing the feasibility of the collected requirements. These three constraints are very critical to RE process for small software organizations. We will discuss these issues in the following case study. As one of succeeding activities, the moderator uses the refined requirements towards further analysis and design activities of software development.

## 4. Case Study

The *Technologies Ahead* [18] is a small software development company in Pakistan which develops applications for mobile phones, has applied our SNS-based RE approach. The company is comprised of 6 employees. It is one of startup companies, which has some recently graduated students majored in computer science and software engineering. We have applied our approach in the *Technologies Ahead* due to some reasons that are 1) the company does not have enough budget to carry out the RE activity and they usually skip the RE phase, 2) the company size was very small in the perspective of staff members, 3) the company wants to develop mobile applications rapidly for sharing the market, and 4) the staff were not fully skilled about the RE process.

Therefore, this company was perfect to validate our SNS-based RE approach.

### 4.1. *Method*

We presented our approach and asked *Technologies Ahead* to apply it to determine end-user requirements for their mobile app development. Generally, the employees were familiar with Facebook and they were using it for social networking. We briefed RE team about the spirit of SNS-based RE approach that would like to elicit requirements for (disaster management app). The RE lead was selected as a moderator who created a discussion room. The moderator asked his team members to invite their friends to discuss the topic in a Facebook group. The moderator also told them that their friends can also invite other friends who could be potential end-users or who have an interest in the topic. Furthermore, the moderator asked participants to utilize the like functionality of Facebook when the discussion has produced significant needs. Especially, the moderator has background work; splitting the users into an expert category and ordinary-user category. The purpose of user categorization was to find more specific features in a precise manner. The experts are those users who have knowledge on the disaster domain (i.e., tsunami, earthquake, floods, etc.) and the ordinary-users are those who have intentions of just using the disaster management app and all those users who come up with general requirements of disaster management app. The categorization of users makes it easy to identify more mature and distinguishable requirements.

The moderator recommended two weeks to conduct the RE activity, but we were not familiar with the dynamics of SNS-based RE approach. Therefore, we agreed to start immediately and did not limit the timeframe. Regarding additional resource support (financial, tools and human resource), we clearly mentioned to provide it if necessary. During RE process, we observed every phase of SNS-based RE approach and monitored each activity closely to evaluate the results.

### 4.2. *Results*

Overall, three employees were involved and 611 users were invited in this case study to discuss the requirements for a mobile application which helps to rescue people in case of any disaster. The end-users discussed the requirements such that the application should inform to people nearby the disaster. The users also discussed that the application should be able to track the location of lost people due to disaster. Some users suggested that the application should support notification "I am safe" shareable through SNS, text, or email. The users also discussed that the application should provide nearest safe zones (by map) in the case of any flood. The users also demanded that the application should be easy to use. Some users from China demanded that the application should support the Chinese language. It is also discussed that the application should guide step by step by instructions to let the user know what to do even before, during, or after the earthquake, storm, or flood even if no data connectivity.

Out of 611 users, 202 users contributed in our case study- meaning that 33.06% users actively participated (by posting comments and likes). Since we had not limited the timeframe, but we have observed that RE activities have become slow down after 4th day and on the 7th day it was negligible as shown in figure 3. The 202 participating users created 719 posts, which mean an average of 3.55 posts per user has been posted. We categorized the posts by content for quality analysis and found that 68.2% posts (490.3) were talking about new features of disaster management app. Elicitation posts are analyzed regarding the level of abstraction of needs posted.[19] We examined the number of "*likes*" to decide about the priority of that requirement. We also recorded the frequency of requirements to decide about their importance. Thus, 345 candidate requirements were gathered, out of which 16 (4.6%) were nonfunctional requirements, which means 1.70 nonfunctional requirements were posted per active user.

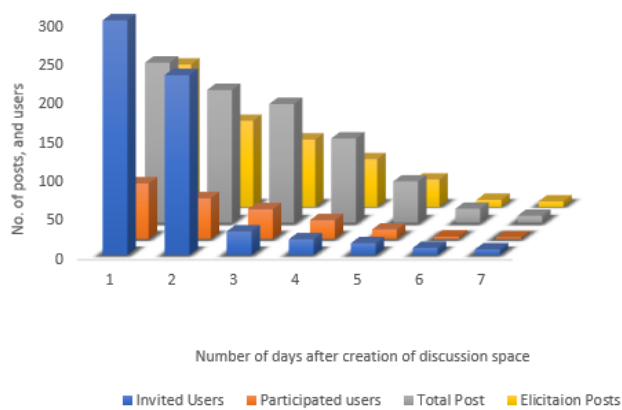

**Figure 3. Participants and Posting for our Requirements Gathering Activity**

### 4.3. Analysis

The requirements prioritization and selection is a very critical activity for small software organizations because the wrong prioritization and selection of requirements not only lead to failure of the project but also becomes a threat to the survivability of software development organization.[20] The small size of development staff, budget constraints, and technical constraints do not allow the small software organization to have a dedicated RE process. Our SNS-based RE approach can cope with financial and technical constraints. For requirements prioritization, Jim Azar[20] proposed an approach for small software development companies. We have used this requirements prioritization technique in requirements analysis phase of our SNS-based RE approach to make it perfect for small software businesses. Table 1 shows the requirements prioritization scheme for disaster management app. Table 1 has three business values and two types of risks. In the matrix, $V_i$ is the weight of business value while $R_j$ is the weight of risk. The requirements prioritization approach then weights each requirement with respect to each specific business value and risk. The weighting scale is 0(not important) to 10(critical).

**Table 1. Requirements Prioritization Matrix**

| Category | Requirements | Core business values | | | Risk | | Scores |
|---|---|---|---|---|---|---|---|
| | | Quality ($V_i=7$) | Effort Required ($V_i=8$) | User Need ($V_i=5$) | Technical ($R_j=-7$) | Business ($R_j=-5$) | |
| Expert | R1 | 4 | 6 | 3 | 5 | 3 | 41 |
| | R12 | 3 | 4 | 4 | 4 | 5 | 20 |
| | R32 | 2 | 5 | 5 | 2 | 4 | 45 |
| | R47 | 6 | 5 | 3 | 7 | 2 | 38 |
| | -- | -- | -- | -- | -- | -- | -- |
| Ordinary-Users | R3 | 6 | 7 | 4 | 6 | 5 | 51 |
| | R10 | 3 | 5 | 3 | 2 | 3 | 47 |
| | R11 | 4 | 6 | 4 | 3 | 3 | 54 |
| | -- | -- | -- | -- | -- | -- | -- |
| | R345 | 5 | 3 | 4 | 5 | 3 | 29 |

The score of each requirement is calculated by the sum of its contribution to core business value minus the sum of its anticipated risk. According to Table 1, the requirement R11 has the highest priority and the requirement R12 has the lowest priority. This priority analysis technique provides a road map for selecting the requirements for further implementation. For small software organizations, the important thing is the feasibility of requirements due to aforementioned constraints. The highest priority of any requirement does not mean that it should be selected. In our case study, the requirement R3 has a higher priority but it is not feasible to implement due to technical constraints and it has been dropped at this stage. During the course of the above process, we filtered futile requirements and dropped unnecessary requirements as well as the user requirements which were not feasible. Therefore, we have identified 156 requirements as the final user requirements of the disaster management app. Out of 156 final requirements, 96 requirements were identified from the users of expert category while 60 requirements were identified from the comments of ordinary-users' category. The final requirements were reaching to 45.22% of the total candidate requirements. The drastic decrease was observed after checking the feasibility of each requirement.

### 4.4. *Findings*

This case study was carried out for the validity of SNS-based RE approach to know how our approach facilitates small software development companies. The main findings of this case study are underlined below:

**End-user Involvement**: The result of this case study shows that 33.06% of total invited users actively participated which shows that our approach can involve anonymous potential users during software development. It has been seen that it was not necessary for end-users to have RE knowledge to participate in our proposed SNS-based RE approach. This characteristic attracts maximum users to be involved during requirements elicitation process. By investigating the ideas and opinions communicated by users, we were able to figure out a high number of user requirements.

**Budget Constraint**: It was observed during case study that additional resources were not needed to carry out requirements elicitation process. Along with other benefits of our approach, the management of *Technology Ahead* was also satisfied with regarding financial matters as well. Therefore, the SNS-based approach was most suitable for *Technologies Ahead.*

**Requirements Elicitation Time**: Figure 3 shows that the moderators invited users only at the beginning (two days) of discussion and our analysis revealed that most requirements elicitation and negotiations posts were created within four days of discussion. We conclude that users were active in the discussion group for a limited timeframe and expressed their needs and concerns within 7 days.

**Knowledge and Skill Constraint**: The moderator was a fresh Computer Science graduate but he has adopted the role of moderator more effectively and gathered 345 user requirements. We have observed that during the requirements prioritization and selection, we have guided the moderator to carry out the prioritization activities.

Therefore, the moderator can learn about RE process but he needs a little bit training to carry RE activities in a smooth way. On the user's part, the knowledge about disaster app has shown its distinction. We have extracted 61.54% of final requirements from the expert users' category and 38.46% of final requirements were collected from ordinary-users' category. We have seen that the comments of expert-users were more logical and realistic than the ordinary-users.

**Non-Functional Requirements Identification**: The result shows that nonfunctional requirements were only 4.6% of total gathered candidate user requirements. Thus, we concluded that end-users only express their needs (business requirements) but do not know much about non-functional requirements.

## 5. Conclusion

The research presented in this paper, propose an SNS-based RE approach that allows a small software company to gather the distributed users' needs from SNS. This approach supports and allows the end-users to participate actively in RE activities such as requirements elicitation, negotiation, and prioritization. Our approach needs minimal training for employees to conduct the RE activity and its cost is almost negligible. Therefore, applying SNS-based approach within SNS make a distinct side to the RE process. We would like to emphasize that the presented RE approach do not necessarily provide well-specified user requirements but rather it reveals end-user needs and supports small companies to identify the initial user requirements within their domain without having any learning overhead to end-users. Although our approach can envision that SNS-based RE approach will become supplementary means for RE activities, and we are confident that it will complement the traditional RE approaches to support small software businesses with limited budget.

Future work is needed to outline the scope, capability, and limitations of such lightweight RE approach. Especially we will focus on how to determine the priority for the collected requirements more precisely. The priority of users' requirements can be different with the project characteristics, for example, project timeline, applied process model, and development team size. Therefore, the priority scheme for requirements should support such affecting factors to elicit more concise users' requirements from SNS.

## 6. Acknowledgment

This research was supported by the National Research Foundation Korea funded by the Ministry of Science, ICT and Future Planning (MSIP), Korea (NRF-2014M3C4A 7030505). We would like to thank the *Technology Ahead* team especially Muhammad Usman Siddique for his support and cooperation.